\documentclass[eqsecnum,aps,prl,twocolumn,showpacs]{revtex4}
\usepackage{graphicx}

\begin{document}
\title{Pressure-induced hole doping in the Hg based cuprates}
\author{C. Ambrosch-Draxl}
\affiliation{Institut f\"ur Theoretische Physik, Universit\"at Graz, Universit\"atsplatz 5, 8020 Graz, Austria}
\author{E. Ya. Sherman}
\affiliation{Institut f\"ur Theoretische Physik, Universit\"at Graz, Universit\"atsplatz 5, 8020 Graz, Austria}
\author{H. Auer} 
\affiliation{Institut f\"ur Theoretische Physik, Universit\"at Graz, Universit\"atsplatz 5, 8020 Graz, Austria}
\author{T. Thonhauser}
\affiliation{Department of Physics, Pennsylvania State University,  University Park, Pennsylvania 16802, USA}


\date{\today}

\begin{abstract}
We investigate the electronic structure and the hole content in the copper-oxygen planes 
of Hg based high $T_c$ cuprates for one to four CuO$_2$ layers and  
 hydrostatic pressures up to 15 GPa. We find that with the pressure-induced 
 additional number of holes of the order of 0.05$e$ the density of states 
 at the Fermi level changes approximately by a factor of 2. At the same time the saddle 
 point is moved to the Fermi level accompanied by an enhanced $k_z$ dispersion. 
 This finding explains the pressure behavior of $T_c$ and leads to the conclusion 
 that the applicability of the van Hove scenario is restricted. By comparison 
 with experiment, we estimate the coupling constant to be of the order of 1, 
 ruling out the weak coupling limit.
\end{abstract}

\maketitle
In the high temperature superconductors, the critical temperature strongly depends on the number $n$ of CuO$_2$ layers per unit cell. In particular, an increase of the transition temperature with increasing $n$ has been verified within the different cuprate families like the Bi, Tl, or Hg based compounds. Although it seemed tempting to add more and more such CuO$_2$ layers to the crystal structure, there is a limit in that for $n>3$ the superconducting properties become worse, i.e. $T_c$ decreases again. Besides the composition, pressure is an important tool to enhance the transition temperature. This effect is most pronounced for the Hg based cuprates \cite{putilin}, which still hold the world record in T$_c$ being 97~K for $n$=1 and ambient pressure, and even 164~K for $n$=3 at 30~GPa \cite{gao94, chu}. Therefore, this system is best suited to simultaneously study both effects.

In this Letter, we report on the electronic properties of HgBa$_2$Ca$_{n-1}$Cu$_n$O$_{2n+2}$ as a function of composition, i.e. $n=1,2,3,4$, and 
hydrostatic pressure, investigated by first-principles calculations. Special emphasis is lead to the question how the charge carriers are redistributed leading to an increase of the hole concentration in the CuO$_2$ layers, which are responsible for the superconducting current. At the same time  the density of states (DOS) at the Fermi level is considered to shed  light on the role of the van Hove singularity (vHS) \cite{bok:87} for the superconductivity and to estimate the coupling constant.

All calculations, which are based on density functional theory (DFT), 
have been carried out with the full-potential linearized augmented planewave (LAPW) 
method utilizing the WIEN2k code \cite{WIEN2k}. The atomic-like basis set used in the atomic spheres around the nuclear positions \cite{singh94,cad:lapw} allows for an analysis of the electronic charge in terms of atomic origin, and for a decomposition of the charge according to its orbital momentum $l$. Therefore,  the decrease of the corresponding occupation number can be interpreted as the creation of a certain amount of holes of the particular character. Below, the CuO$_2$ layers will be analyzed that way. 
The most important orbitals are Cu-$d_{x^2-y^2}$ and O-$p_x$ forming $pd-\sigma$ bonds. 
Note that the O-$p_y$ orbital of the second oxygen neighbor is equivalent to O-$p_x$ 
orbital of the other oxygen due to symmetry.
For all our calculations, the crystal structures have been relaxed in terms of lattice 
parameters and atomic positions as determined and discussed in detail in Ref. \onlinecite{timo:03}. 
This procedure makes the calculations not to rely on experimental crystalline data. 
The latter is important for studying the pure dependence of the electronic properties
on pressure or $n$, since the measured samples usually include some excess oxygen and
can be strongly Hg deficient \cite{antipov02}. 
\begin{figure}
\begin{center}
\includegraphics[width=\columnwidth]{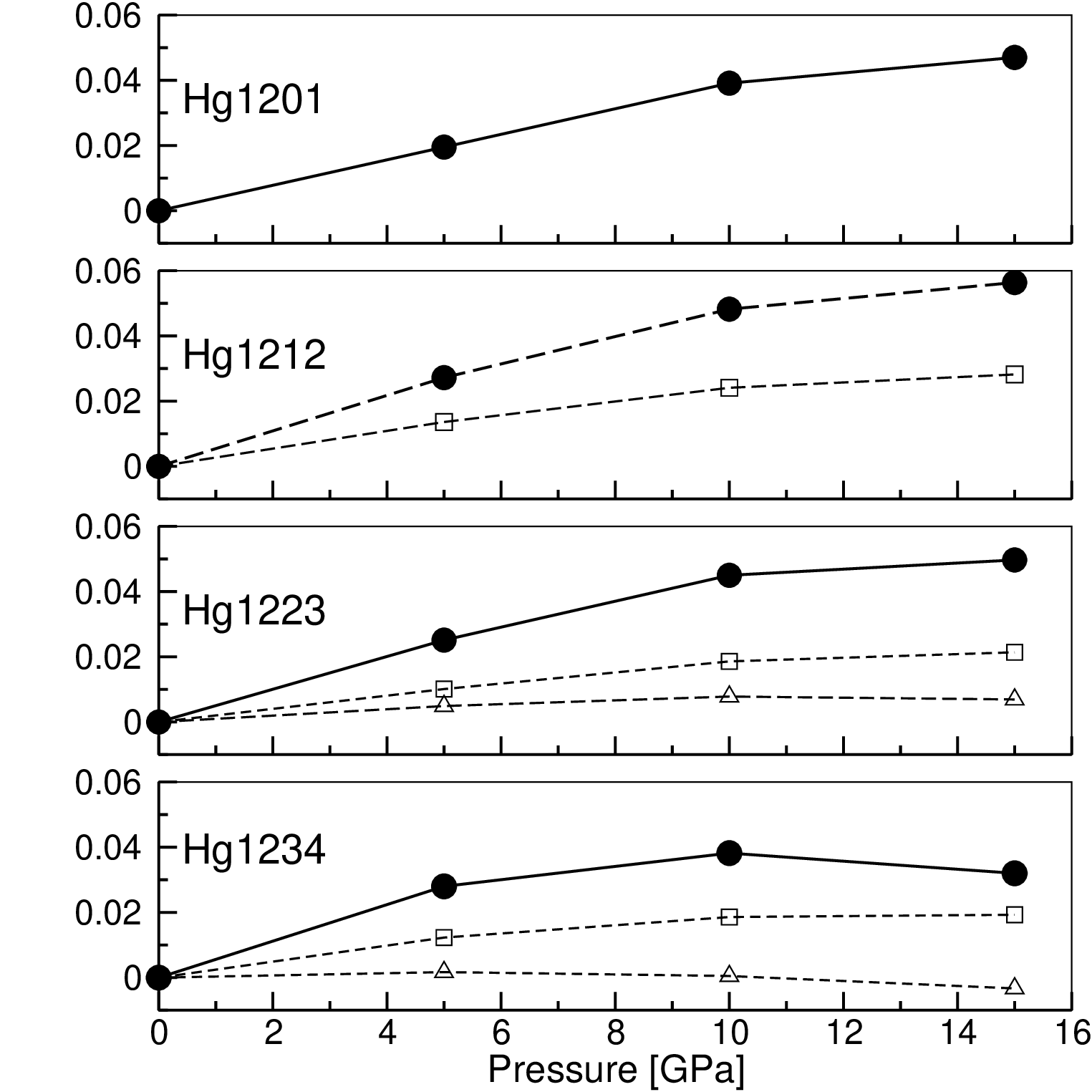}
\end{center}
\caption{\label{fig:holes_P}Number of pressure-induced holes (in units of electrons) 
for the Hg based cuprates, Hg1201, Hg1212, Hg1223, and Hg1234 for pressures up to 15~GPa. 
The values are taken with respect to ambient pressure for each material.
The filled circles correspond to the total number of holes created per 
unit cell, whereas the open symbols (triangles for innermost and squares for outer layers)
reflect the situation in a particular layer. 
The symbols indicate the calculated data points, while the lines serve as guide to the eye.}
\end{figure}
Figure \ref{fig:holes_P} depicts the number of holes created in the Cu-$d_{x^2-y^2}$ orbitals when 
pressure is applied. Before discussing the details, it should be noted that for all compounds 
and pressures the atomic sphere radii, chosen as 2.1, 2.3, 1.9, 1.2, and 2.0 a.u. for Hg, Ba, Cu, O, and Ca, 
respectively, in which the electronic charge is analyzed, has been kept constant. 
This means that a bigger  fraction of the crystal volume and hence electronic 
charge is lying inside the sphere when pressure is applied / increased. 
Therefore, the number of pressure-induced holes can be regarded to be somewhat 
underestimated by our pressure calculations. Nevertheless, it can be stated that 
there is a pronounced pressure effect on the orbital occupation numbers of 
copper, whereas there is hardly any changes in the corresponding oxygen 
O-$p_x$ charges \cite{timo:03}. This is in contrast to the  
doping-induced holes \cite{cad:03a} which are found not only at 
the copper sites, but also, to some extent on the oxygen positions as well.

Let us first focus on the partial charges at ambient pressure. When changing the composition, 
the Cu-$d_{x^2-y^2}$ occupation number changes from 1.4336 for $n=1$ to 1.4414 for $n=2$, 1.4421 (1.4386) for $n=3$, and 1.4413  (1.4380) for $n=4$. The numbers in brackets refer to the inequivalent copper positions occurring for the three and four layer compounds and indicate an inhomogeneous hole distribution 
that concerns the different CuO$_2$ layers within one unit cell. 
Upon applying pressure, the number of holes increases with a parabolic-like shape which 
at 15~GPa is close to its maximum for $n$ up to three. Only for $n$=4, there is already 
a slight decrease in the highest pressure range. An important fact is that the 
amount of pressure-induced charges per site decreases for higher $n$. 
In particular, the innermost layers (marked by triangles 
in Fig. \ref{fig:holes_P}) do not gain more holes with pressure 
beyond $n=3$. Only the total amount of holes per cell created for a 
certain pressure value is similar in all the compounds, i.e. when the 
contributions from different layers are summed up weighted by their multiplicity. 

All these findings are fully concomitant with the superconducting
transition temperature \cite{antipov02}: Experimentally, the enhancement of 
$T_c$ for $n$ up to 3 with pressure is approximately 15-20 K between 0 and 15 GPa, 
i.e. it is nearly independent of $n$. Only for $n$=4 the curve is somewhat flatter. 
Exactly the same trend is found for the number of pressure-induced holes. 
The inhomogeneous charge distribution among the different layers also provides 
an explanation for the limited $T_c$ since the inner layers do not gain any 
more holes by further insertion of CuO$_2$ planes.
 
The fact that  $T_c$ scales with the number of charge carriers does, however, 
not provide an explanation for the origin of superconductivity. Although the 
amount of holes in the cuprate planes is an important quantity, its knowledge 
does neither lead to a conclusion about the pairing mechanism, nor about the 
coupling strength to the mediating quasiparticles. Therefore, the question 
arises what the role of the bands and the density of states around the Fermi 
level is. In this context, the van Hove scenario has been frequently invoked. 
Indeed, the most precise calculations for the bandstructure of 
YBa$_2$Cu$_3$O$_7$ \cite{cad:99} exhibit a saddle point (usually referred to as a vHS) 
pinned to the Fermi level, and the same is 
true for La$_{2-x}$Ba$_x$CuO$_4$ \cite{timo:02} and HgBa$_2$CuO$_{4+\delta}$ at 
optimal doping \cite{cad:03a}. 

\begin{figure}
\begin{center}
\includegraphics[width=\columnwidth]{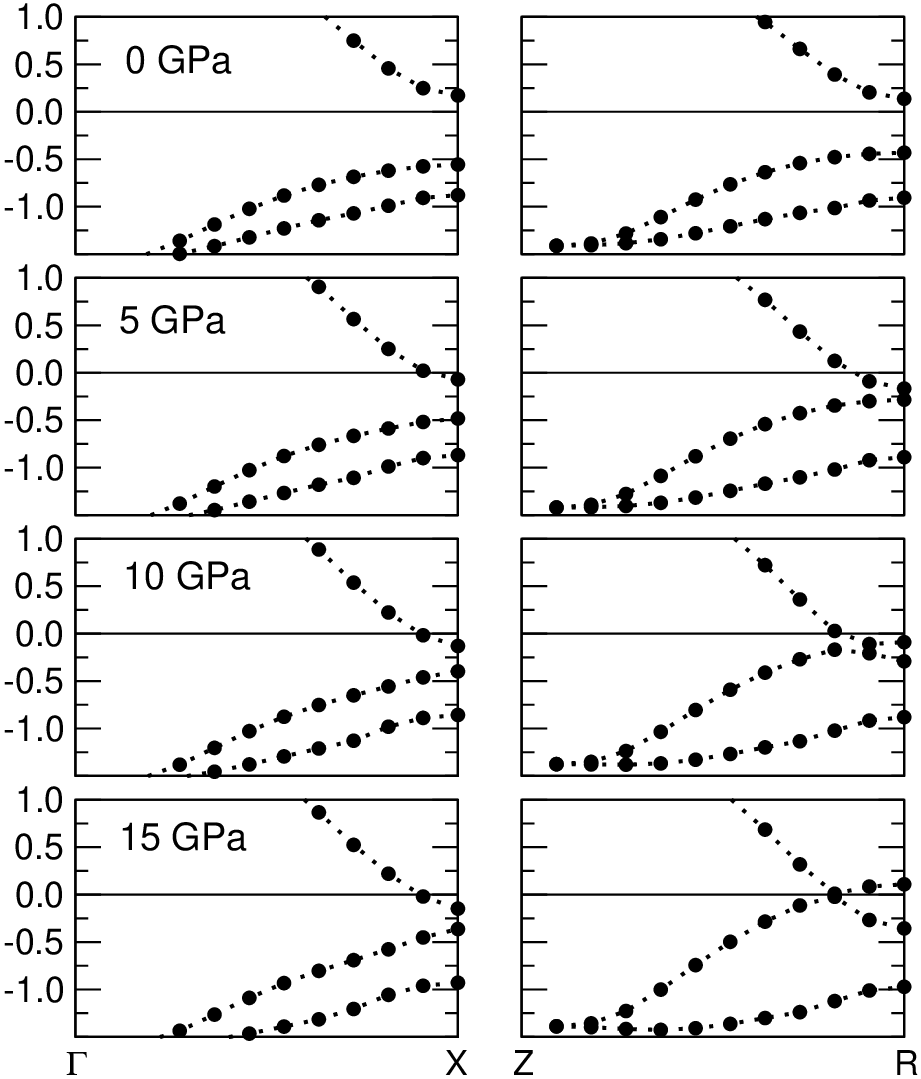}
\end{center}
\caption{\label{fig:Hg1201_bs}Band structure of Hg1201 along the high symmetry 
lines $\Gamma X$ and $ZR$ for different pressure values. The black filled circles 
mark the calculated points, where the dotted lines serve as guide to the eye. 
The Fermi level is set to zero and indicated by a horizontal line, and the energy is given in eV.}
\label{fig:bands}
\end{figure}

Here, we will assess this possibility for the Hg based compounds under pressure 
by performing selfconsistent calculations for all pressure values to obtain the 
corresponding electronic structures. Thereby we observe two effects: 
First, upon applying pressure, the dispersion in the $k_z$ direction 
is strongly increased.  The band width of the  $pd-\sigma$ band along $XR$ is 
changed from 0.12 to 0.29 eV between 0 and 10 GPa which corresponds to a factor of 2.5. 
Second, at the same time the ratio of hopping matrix elements $t'/t$ decreases 
by 6\%, where $t'$ and $t$ denote the in-plane matrix elements of O-O and 
Cu-O hopping when mapping the band structure to a simple tight binding model.  
These two effects lead to an asymmetry in the location of the 
vHS with respect to the Fermi level for different $k_z$ values.  
This can be seen in Fig. \ref{fig:Hg1201_bs}, where the pressure 
dependence of the Hg1201 band structure around $E_F$ along the two high 
symmetry lines $\Gamma$$X$ and $ZR$ is depicted. Upon applying pressure, 
the shoulder between $Z$ and $R$ clearly moves up toward the Fermi level 
intersecting it at around 10 GPa. Along  $\Gamma$$X$ the band is pushed 
up as well thereby getting steeper, but staying below $E_F$ in the entire 
pressure range. As a result, the vHS is broadened when it arrives at the Fermi level.
Thus, we have found the vHS to be pinned to the Fermi level within a certain region 
of the Brillouin zone (between $\Gamma$$X$ and $ZR$) at high pressures which could be 
related to the high and rather weakly pressure-dependent $T_c$ in a pressure range between 10 and 20 GPa. 

\begin{figure}
\begin{center}
\includegraphics[width=\columnwidth]{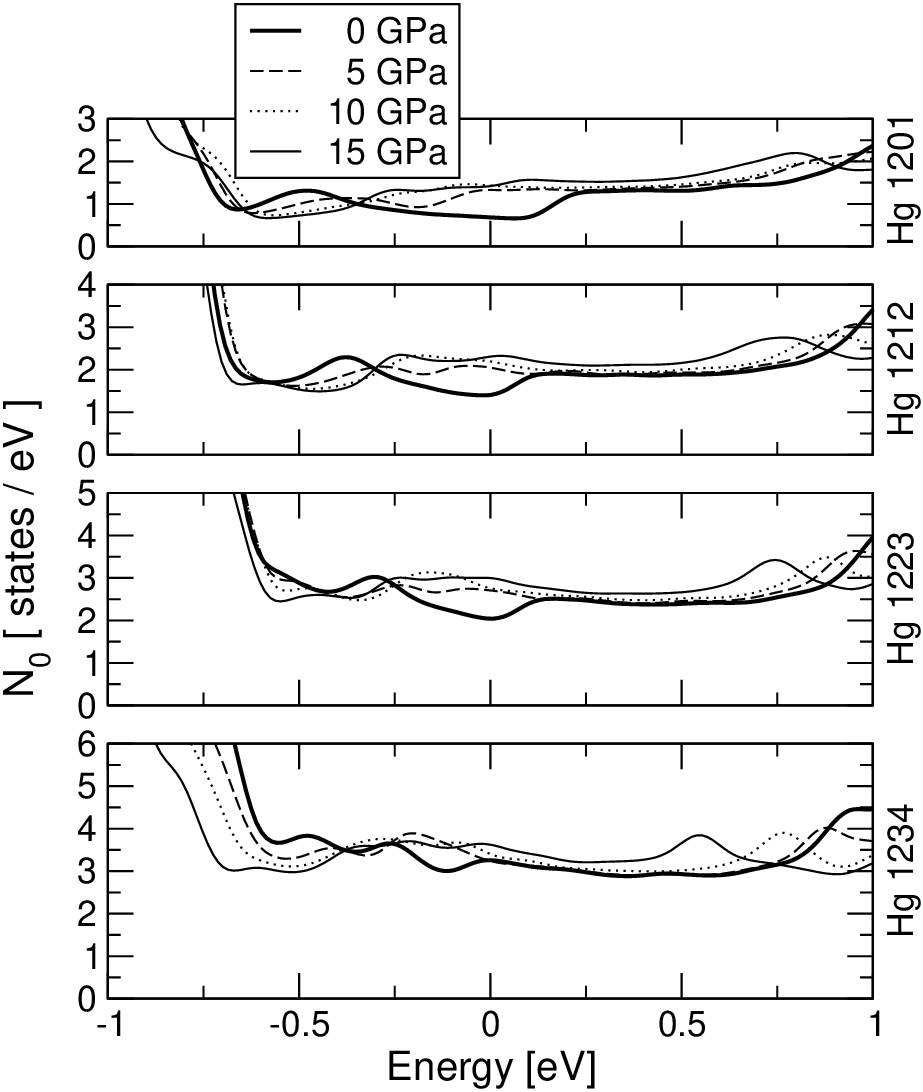}
\end{center}
\caption{\label{fig:dos}Densities of states $N$ in states per eV and unit cell of Hg1201, Hg1212, Hg1223, and Hg1234 for pressure values of  0, 5, 10 and 15~GPa.}
\end{figure}

\begin{figure}
\begin{center}
\includegraphics[width=\columnwidth]{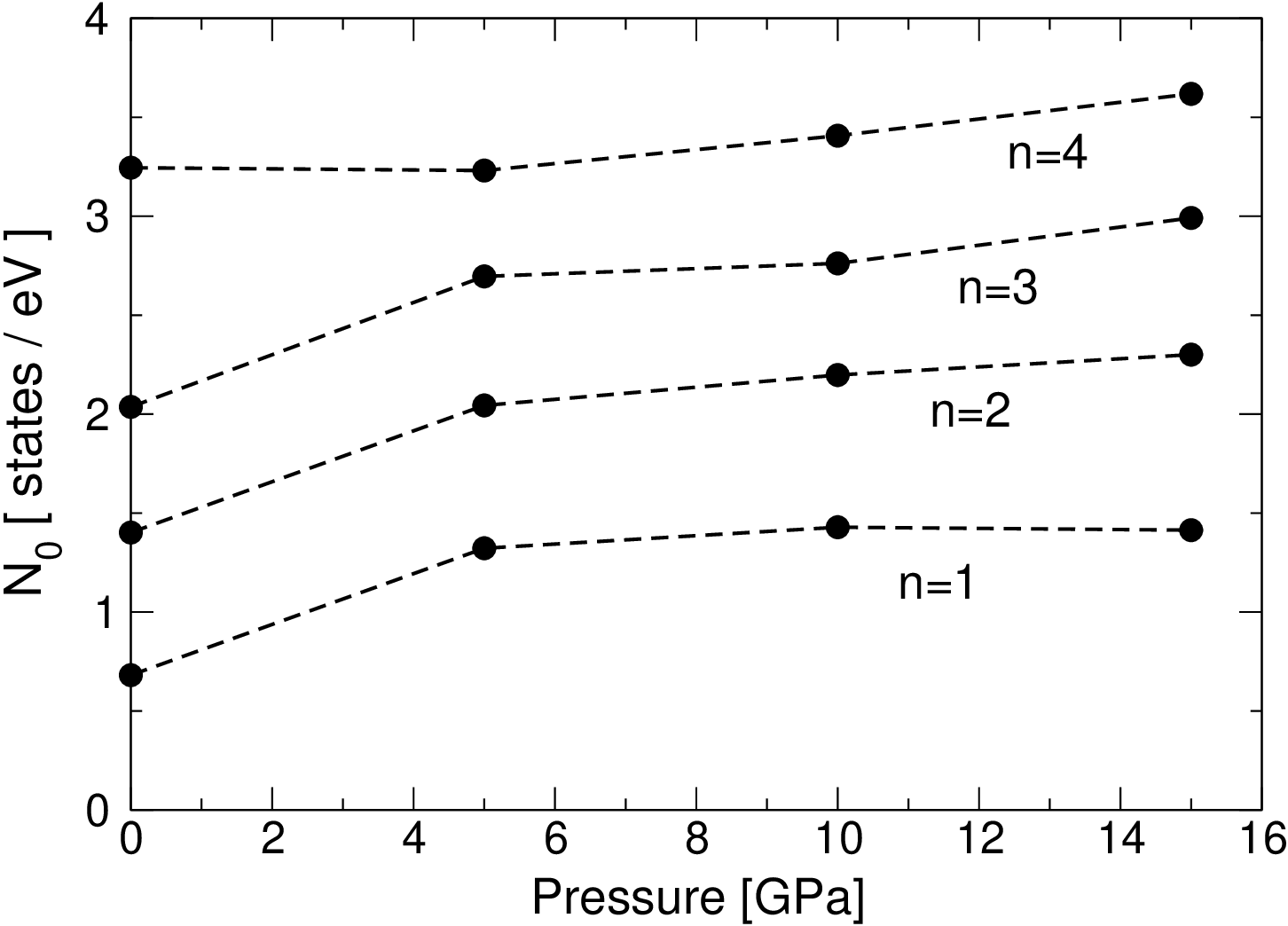}
\end{center}
\caption{\label{fig:dos_0}Densities of states at the Fermi level $N_0$ in states per eV and unit cell in Hg1201, Hg1212, Hg1223, and Hg1234 for pressure values of  0, 5, 10 and 15~GPa.}
\end{figure}

These findings also show up in the density of states depicted in Fig.  \ref{fig:dos}. 
For Hg1201, the shoulder in the DOS is located at -0.5 eV but moving up toward $E_F$ with pressure, 
where the density of states at the Fermi level  $N_0$ is only slowly varying for higher 
pressure values (see Fig. \ref{fig:dos_0}). For $n$=3, the situation is the same, while 
for $n$=2 and 4 the vHS crosses $E_F$ between $ZR$ and stays below $E_F$ along $\Gamma$$X$. 
Concerning the magnitude of the DOS the following is interesting to note: $N_0$ is 
getting larger with the number of layers, where the pressure-induced increase is 
nearly the same for  $n$=1, 2, and 3. The reason for the latter fact is that with 
additional CuO$_2$ layers the corresponding bands are not fully degenerate but exhibit 
a certain splitting which does not allow all the corresponding shoulders (vHSs) to sit 
at $E_F$ at the same time. Thus, the pressure-effect on the DOS is independent of the 
number of layers, at least up to  $n$=3. This trend goes hand in hand with the number of 
holes as well as with the critical temperature. Only for $n$=4, the pressure effect is 
less pronounced, since at ambient pressure, $N_0$ does not fall into a local minimum 
as is the case for the other compounds. 

We want to point out that the selfconsistent treatment is crucial for understanding 
the changes of the band structure under pressure. For example, when using the rigid 
band approximation to estimate the pressure-induced number of holes necessary to push 
the vHS to the Fermi level, as it was proposed first by Novikov {\it et al.} \cite{novikov}, 
this quantity being of the order of 0.5$e$ is overestimated by about an order of magnitude (see Fig \ref{fig:bands}). 

Knowing on the one hand, the pressure dependence of $N_0$ for the four compounds as a
 result of our calculations, and the corresponding experimentally measured $T_c$s 
 on the other hand, one can try to estimate the effective dimensionless coupling 
 constant $\lambda=VN_0$, assuming a BCS-like behavior valid for the weak coupling 
 for any type of the intermediate boson, $T_c \sim \omega_B \exp(-1/\lambda)$. 
 Here,  $\omega_B$ is the frequency of the mediating boson, and $V$ characterizes the 
 coupling strength. It should be noted here that the DOS does not necessarily have to be 
 assumed to be constant near the Fermi level on the $\omega_B$ scale, but it turned out 
 to be approximately the case (Fig. \ref{fig:dos}) and thus can be used for such an estimation. 
 There are, however, several problems which hamper a quantitative comparison between theory and experiment. 
(i) The main concern comes from the fact that $T_c$ has been measured on samples which 
include a doping concentration close to optimal doping. This doping could be considered as an 
unknown chemical pressure. On the contrary, the calculations have been performed for the 
undoped material.
(ii) There are big uncertainties in the experimental data regarding defects and non-stoichiometry 
of the samples. 
(iii) When dealing with compounds with more than one layers, it is unclear how their 
contributions enter $T_c$. 

Despite all these uncertainties, conclusions are nevertheless possible. First, from all 
calculations performed to investigate doping as mentioned before for YBa$_2$Cu$_3$O$_7$ \cite{cad:99},  
La$_{2-x}$Ba$_x$CuO$_4$ \cite{timo:02} and HgBa$_2$CuO$_{4+\delta}$ \cite{cad:03a} we know that at 
optimal doping the saddle point is  pinned to the Fermi level. Since in Hg1201 
the vHS reaches $E_F$ at approximately 5GPa, we interpret this value as the effective "chemical" 
pressure corresponding to optimal doping in the sense that both the optimal doping and this pressure 
lead to similar changes in the electronic bandstructure in the 
vicinity of the Fermi level, which are important for superconductivity.   The changes of $T_c$ due to pressure occur on top of the 
doping-induced $T_c$, i.e. are of the order of 10-20\%. The same is observed in $N_0$ when 
increasing pressure above 5 GPa as shown in Fig. \ref{fig:dos_0}. From a comparison of the 
pressure-induced change of $T_c$ with the change of $N_0$ we  can estimate $\lambda = \Delta N_0 / \Delta T_c \times T_c/N_0$ being of the order of one. Taking into account a possible pressure enhancement of $\omega_B$ can only increase this estimated value. Therefore, we can rule out the BCS-like weak coupling approaches. 

Second, we want to comment on the van Hove scenario. We find that it
is not the driving force for boosting $T_c$ in the sense that it never shows 
up as a sharp peak in the density of states at the Fermi level. At ambient pressure, 
there exists only a saddle point in the band structure, i.e. a rather broad feature in 
the density of states which is further considerably broadened when pressure is applied.  
Thus, the effect of a non-uniform density of states near the Fermi level, which was supposed 
to be important to enhance $T_c$ \cite{Horbach} in comparison with standard BCS theory, is 
not applicable in this case. Nevertheless, this saddle point plays an important role as 
it is pinned to the Fermi level at optimal doping. The pressure effect on $T_c$ can be 
traced back to the above discussed pressure-induced $z$ dispersion and the corresponding 
changes in the hopping matrix elements.

Summarizing our results, we find the hole content in the CuO$_2$ planes to increase as a 
function of pressure by approximately 0.05 electrons going from ambient pressure to 15 GPa. 
These changes in the hole concentration are accompanied by a considerable gain in the density of states 
at the Fermi level arising due to the movement of the saddle point up to and even 
through the Fermi level. The latter effect is caused by an increase in the $z$ axis 
dispersion due to pressure. We find the effect of optimal doping to be equivalent 
to applying a pressure of approximately 5GPa. On top of this, the increase in the 
density of states is of the order of 10\% in parallel to the enhancement of $T_c$ which is 
of the same order of magnitude. From this comparison we conclude that superconductivity 
in this cuprate family is described by the intermediate to moderately strong coupling regime, 
i.e. $\lambda$ being of the order of 1. The saddle point, often referred to as a vHS, 
does not show up as a peak in the density of states making models based on a non-uniform 
density of states being not applicable to high $T_c$ cuprates.

\vspace{0.5cm}
\noindent
{\bf Acknowledgments} \newline
We appreciate support from the Austrian Science Fund, projects P13430, P14004, and M591, 
and by NSF Grant No.\ DMR-02-05125. Most of the calculations were done at the Materials 
Simulation Center, a Penn-State MRSEC and MRI facility. C.A.D thanks E. Antipov for valuable discussions.

\end{document}